\documentclass[preprint,review,12pt]{elsarticle}

\usepackage{graphics}
\usepackage{epsfig} 
\usepackage{amsmath}
\usepackage{amssymb}
\usepackage{fancyhdr}
\usepackage[a4paper]{geometry}

\begin{document}
\pagestyle{fancy}
\lhead{}
\chead{}
\rhead{\textbf{\large P1-66}}
\begin{center}
\renewcommand{\baselinestretch}{2.0}

\large{\textbf{Numerical investigation of Scrape Off Layer anomalous particle transport for MAST parameters}}

\vskip 6pt
\normalsize{\textbf{F. Militello}$^{a,*}$, W. Fundamenski$^{a,b}$, V. Naulin$^c$, A.H. Nielsen$^c$}

\vskip 6pt
\textit{
$^a$ EURATOM/CCFE Fusion Association Culham Science Centre, Abingdon, UK. \\
$^b$ Imperial College of Science, Technology and Medicine, London, UK.\\
$^c$ Association EURATOM-DTU, Technical University of Denmark, Roskilde, Denmark.}
\end{center}

\renewcommand{\headrulewidth}{0.0pt}
\renewcommand{\baselinestretch}{2.0}
\normalsize
\section*{Abstract}
\noindent
Numerical simulations of L-mode plasma turbulence in the Scrape Off Layer of MAST are presented. Relevant features of the boundary plasma, such as the thickness of the density layer or the statistics of the fluctuations are related to the edge density and temperature, plasma current and parallel connection length. It is found that the density profile is weakly affected by the edge density, it broadens when the current or the temperature are decreased while the connection length has the opposite effect. The statistics of the turbulence is relatively insensitive to variations of all the edge parameters and show a certain degree of universality. Effective transport coefficients are calculated for several plasma conditions and display a strong nonlinear dependence on the parameters and on the radial variable. Finally, it is shown how the perpendicular particle fluxes in the Scrape Off Layer are related to the edge parameters. 

\normalsize
\noindent \textit{PACS}:\\
\noindent \textit{JNM keywords}: Theory and Modelling\\
\noindent \textit{PSI-20 keywords}: Edge modelling, Fluctuations and turbulence\\
\noindent$^{*}$\textit{Corresponding Author Address}: EURATOM/CCFE Fusion Association Culham Science Centre, Abingdon, Oxon, OX14 3DB, UK\\
\noindent$^{*}$\textit{Corresponding Author E-mail}: fulvio.militello@ccfe.ac.uk\\
\noindent \textit{Presenting Author}: Fulvio Militello\\
\noindent \textit{Presenting Author E-mail}: fulvio.militello@ccfe.ac.uk\\

\pagebreak

\renewcommand{\baselinestretch}{2.0}
\normalsize

\section{Introduction}

The heating and fusion power in the next generation of tokamaks will be much larger than in present-day experiments. The exhaust of this power toward the solid surfaces of the machine needs to be controlled and limited in order to avoid unacceptable material degradation \cite{Loarte2007}. At the same time, the perpendicular particle transport determines where the interactions between plasma and solid surfaces occurs \cite{LaBombard2000}. Crucially, both the power and particle deposition are governed by the physics of the narrow region surrounding the hot, well confined plasma inside the ideal separatrix. 

It was soon realized that the plasma in the Scrape-Off Layer (SOL) has a peculiar behaviour, quite different from the core \cite{Zweben1985,Wootton1990}. Open field lines connected to solid surfaces and the presence of neutrals and impurities coming from the wall characterise this relatively cold and rarefied environment. The result is a plasma far from equilibrium with large fluctuations and an intermittent behaviour \cite{Zweben1985,Boedo2001}. 

Most of the numerical studies previously carried out to characterise the SOL focused on the ability of the codes to match single experimental observation or to reproduce the special features of the turbulence in this special region of the machine (i.e. intermittency, non Gaussian behaviour, large fluctuations). The ESEL code was particularly successful in both these applications \cite{Garcia2004,Garcia2005,Garcia2006,Fundamenski2007,Militello2012a}. Based on these results, we used ESEL with an alternative approach, which we followed in \cite{Militello2012b} and in the present work, and that consists in exploring the operationally achievable parameter space of a machine and in determining how the SOL responds to variations of edge conditions (e.g. density, temperature and current). As compared to experimental scans, which can only be done by varying engineering quantities (fuelling, input power, total plasma current), our numerical scans have the benefit of clearly separating the effect of a specific edge quantity from the others (which are kept constant during the scan). 

In this contribution, we focus on how edge temperature, density, current and connection length affect the thickness of the SOL (measured by the \textit{density} decay length), the effective diffusion and velocity, and the particle fluxes. In other words, we study the anomalous particle transport in the SOL as the conditions in the plasma boundary are varied. A longer discussion on the density and tempartaure profiles as well as the statistical properties of the fluctuations can be found in \cite{Militello2012b}. Our investigation is based on numerical simulations of L-mode plasma regimes relevant for the Mega Ampere Spherical Tokamak (MAST), selected experimental data of which were successfully reproduced by the ESEL code \cite{Militello2012a}.  

\section{Model and numerical set up}

The model we solve is described by a set of drift-fluid equations for the evolution of the plasma density, $n$, electron temperature, $T$, and plasma vorticity, $\Omega$. The latter is related to the electrostatic potential, $\phi$, as the perpendicular plasma velocity is identified with the $\textbf{E}\times\textbf{B}$ drift. The equations are:
\begin{eqnarray}
\label{1}
\frac{\partial n}{\partial t} +\frac{1}{B}[\phi,n] &=& nC( \phi) - C(nT_e) +D\nabla_\perp^2n -\Sigma_n n, \\
\label{2}
\frac{\partial T}{\partial t} +\frac{1}{B}[\phi,T] &=& \frac{2}{3}TC(\phi) - \frac{7}{3}TC(T)-\frac{2}{3}\frac{T^2}{n}C(n) +\chi\nabla_\perp^2T -\Sigma_T T, \\
\label{3}
\frac{\partial \Omega}{\partial t} +\frac{1}{B}[\phi,\Omega] &=& -C(nT_e)+\mu\nabla_\perp^2 \Omega -\Sigma_\Omega \Omega, \\
\label{4}
\Omega &=& \nabla_\perp^2 \phi.
\end{eqnarray}  
Here, $B^{-1}[\phi,f]\equiv B^{-1}\textbf{b}\times\nabla\phi \cdot \nabla f$ (representing the advection of a generic field $f$ due to the $\textbf{E}\times\textbf{B}$ drift). $C(f) \equiv (\rho_s/R_0) \partial f/\partial y$ is the curvature operator, where $R_0$ is the major radius, $\rho_s$ is the ion Larmor radius calculated with the electron temperature and $y$ is the "poloidal" coordinate in the drift plane. The magnetic field, $B$ is assumed to vary as the inverse of the major radius, so that $B^{-1} \approx 1+ \epsilon +(\rho_s/R_0) x$, where $\epsilon$ is the inverse aspect ratio and $x$ is the "radial" coordinate. A Bohm normalization is used to make the equations dimensionless. 

On the right hand side of Eqs.\ref{1}-\ref{3} appear terms that represent the perpendicular collisional dissipation (with particle diffusion, $D$, the heat conductivity, $\chi$, and the ion viscosity, $\mu$) and parallel losses ($\Sigma_n$, $\Sigma_T$ and $\Sigma_\Omega$ are the inverse of the typical loss times). The parameters $D$, $\chi$ and $\mu$ are calculated using neoclassical theory, $\Sigma_n=\Sigma_\Omega$ are obtained assuming advective losses in the parallel direction while conductive losses are used to estimate $\Sigma_T$ \cite{Fundamenski2007,Militello2012b}. More details on the ESEL model and the physics that it describes can be found in \cite{Garcia2004,Fundamenski2007,Militello2012b}. 

We define our reference case with the following parameters $n_0=0.8\cdot 10^{19} m^{-3}$, $T_{0}=40 eV$, $B_0=0.5 T$, $q=7$, $L_\parallel=10 m$, $\epsilon=0.69$, $R_0=0.85 m$ (we used here dimensional values). These quantities uniquely define all the dimensionless parameters in Eqs.\ref{1}-\ref{4} \cite{Fundamenski2007,Militello2012b} and therefore control the physics of the system. In particular, $\rho_s=1.83$ mm for the reference case. During the scans, one of the dimensional parameters is varied while the others are kept at the reference value. 

The simulation domain where Eqs.\ref{1}-\ref{4} are solved covers  $150\rho_s$ in the $x$ direction, $100\rho_s$ in the $y$ direction and is resolved with 512x256 points (we checked that a non-equispaced grid did not affect our calculations). In order to capture the transition between closed to open field lines, the parallel losses are taken to be a function of the radial variable. In the first $50\rho_s$ we have $\Sigma_n=\Sigma_T=\Sigma_\Omega=0$, representing the edge plasma inside the separatrix, while in the remaining $100\rho_s$ the parallel loss coefficients take a finite value determined by the edge parameters. Finally, we fix $T=1$, $n=1$ (i.e. the dimensional temperature and density take the values $T_0$ and $n_0$, respectively) and $\Omega=\phi=0$ at the inner boundary, $\Omega=v_y=0$ and $\partial T/\partial x =\partial n/\partial x=0$ at the outer boundary and we impose periodicity in the "poloidal" direction. All the simulations are run for a few tens of $msec$, which corresponds to several thousands turbulence correlation times (i.e. our results are obtained when the turbulence is in a statistically steady state). Similarly, the coherent turbulent structures are well resolved and much smaller than the domain size.  

\section{Scans in the edge parameters}

\subsection{Edge density}

\begin{figure}[!h]
\centerline{\scalebox{0.8}{\includegraphics[clip]{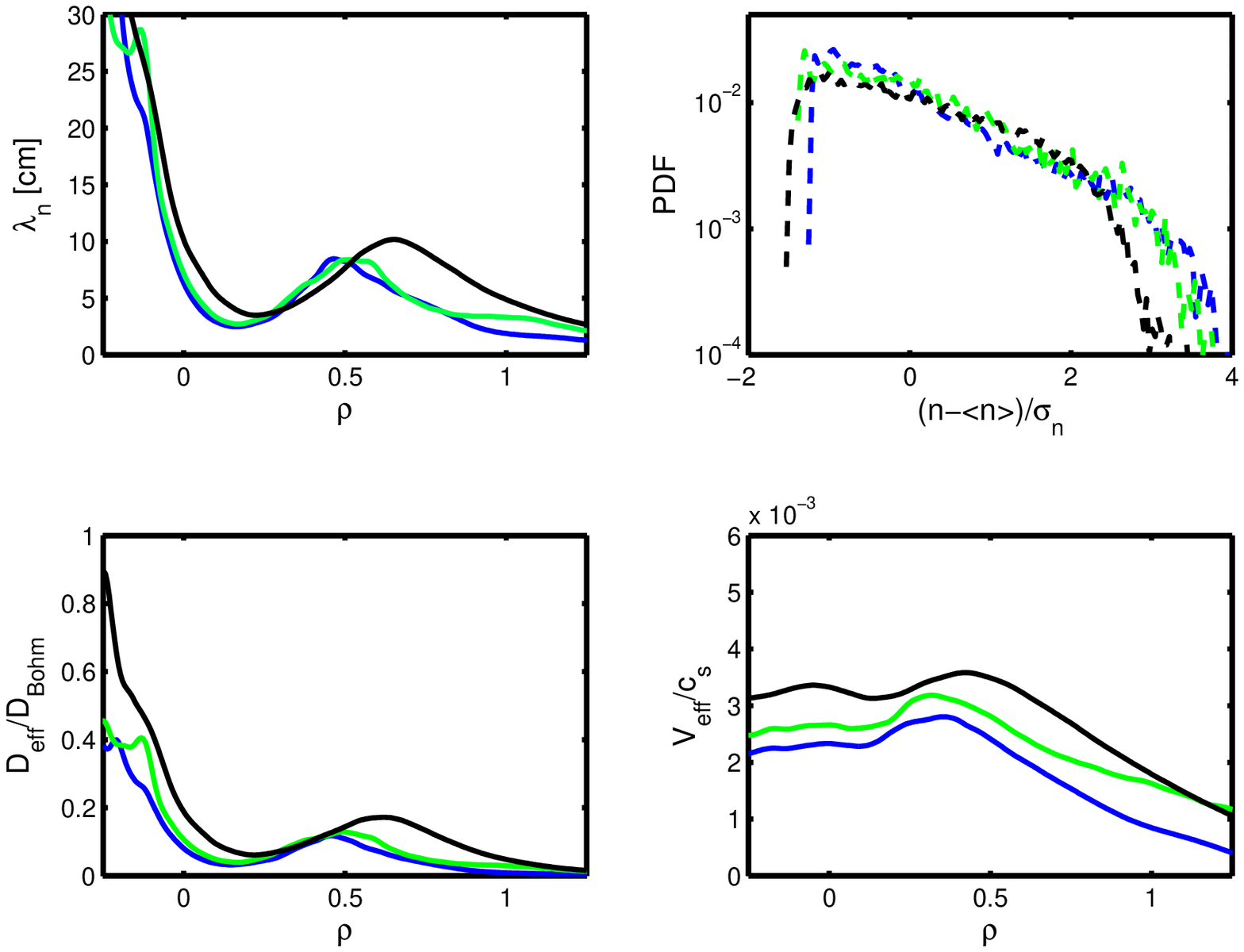}}}
\caption{}
\label{figure1}
\end{figure}

The first scan we discuss is in the edge density. Figure \ref{figure1} gives a description of how this quantity affects some important features of the SOL. In all the frames, the three curves represent $n_0=0.8\cdot 10^{19}$ $m^{-3}$ (blue), $n_0=1\cdot 10^{19}$ $m^{-3}$ (green) and $n_0=1.4\cdot 10^{19}$ $m^{-3}$ (black). 

The upper left frame shows the radial profile of the density decay length, defined as $\lambda_n \equiv -<n>/(d<n>/dr$) where $<n>$ is density profile averaged in time and in the poloidal direction, while $r=\rho_s x$ (so that $\lambda_n$ is in dimensional units). In addition, $\rho\equiv x/50$ is a normalized radial variable. Note that the decay length varies with the radius, which indicates that the average profile of the density is not falling exponentially, as it is sometimes assumed in other modelling works. The upper right frame shows the probability distribution function (PDF) of the density fluctuations evaluated at $\rho=0.4$ (corresponding to $r \approx 3.6$ cm in the SOL). Note that in the figure, $\sigma_n$ is the standard variation of the distribution. The statistics of the fluctuations are clearly not Gaussian, and are characterised by a large number of large positive events, corresponding to blobs of plasma propagating in the SOL (this is a sign of intermittent behaviour). The skewness and flatness of the PDF at $\rho=0.4$ are $1.5$ and $2$, respectively \cite{Militello2012b}.

The lower frames show the effective anomalous diffusion, $D_{eff}/D_{Bohm}\equiv -<\Gamma>/(d<n>/dx)$, and the effective anomalous velocity, $V_{eff}/c_s\equiv <\Gamma>/n$ (note that $<\Gamma>$ is the averaged particle flux). Both quantities are normalized, the former to the Bohm diffusion, $D_{Bohm}\equiv \rho_s c_s$ and the latter to the cold ions sound speed, $c_s\equiv \sqrt{T_0/m_i}$ (where $m_i$ is the ion mass). The usefulness of $D_{eff}$ and $V_{eff}$ relies on the validity of local transport, which is modelled by Fick's law. This paradigm fails in the SOL, where the turbulent transport is non-local and memory effects are important. As shown in the figure, $D_{eff}$ and $V_{eff}$ have significant radial variations (more than a factor 2), which implies that it is not possible to characterise and parametrise the cross field turbulent transport in neither of this simple way nor a combination of them (this is consistent with experimental observations, see \cite{LaBombard2000,Garcia2007}). 

It is interesting to note that in each frame of Fig.\ref{figure1}, the curves do not significantly differ from one another, which suggests that in this regime the SOL does not respond to edge density variations. This, however, does not mean that there is no change when the \textit{line averaged} density is modified, as was indeed observed in several experiments \cite{LaBombard2000,Garcia2007a}. This effect is more likely related to the concurrent variation of the edge temperature which does control the SOL features, as will be discussed in the next subsection. 

\subsection{Edge temperature}

\begin{figure}[!h]
\centerline{\scalebox{0.8}{\includegraphics[clip]{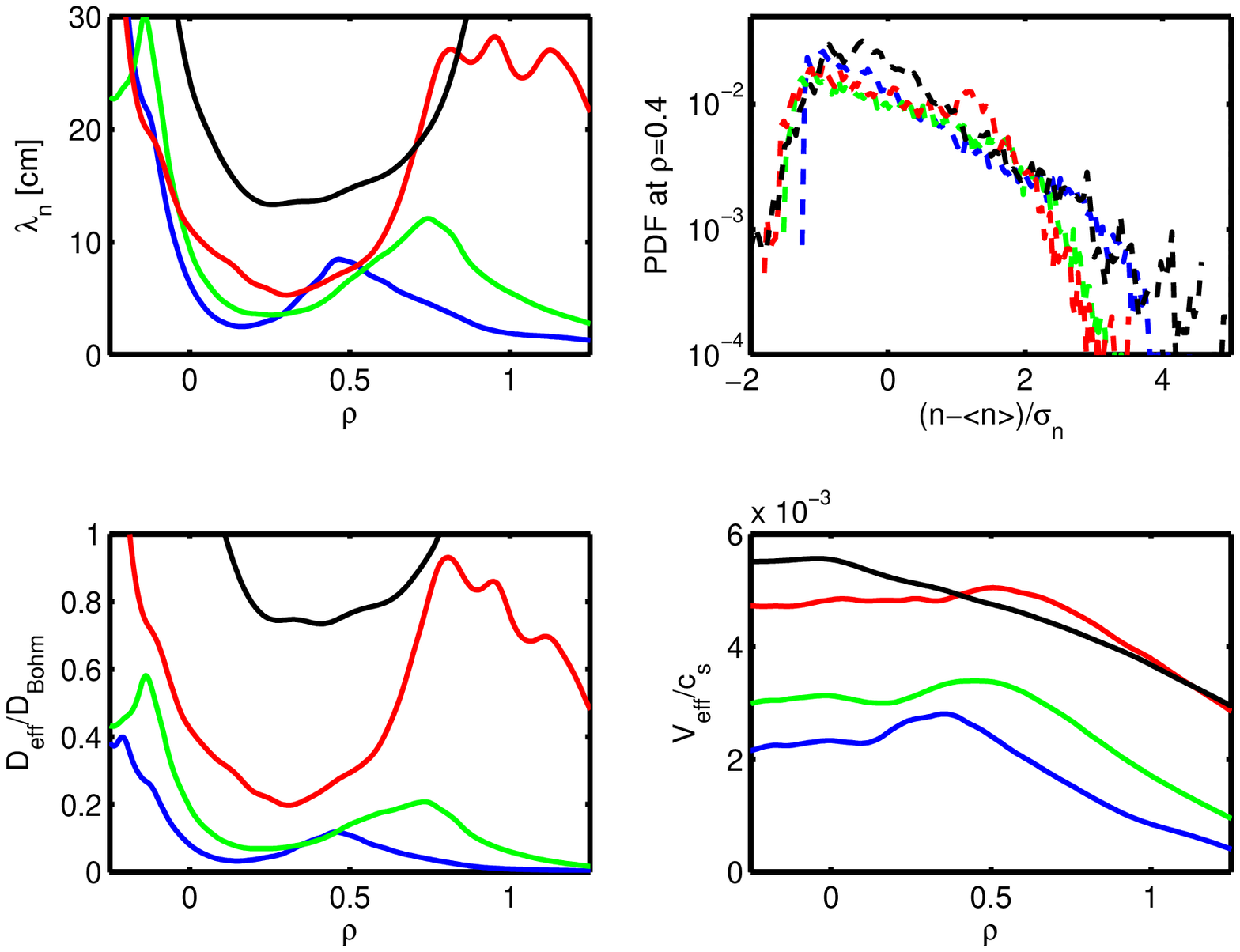}}}
\caption{}
\label{figure2}
\end{figure}

Differently from the edge density, we found that the edge temperature plays a big role in determining the characteristics of the SOL. The scan we performed covers $T_0$ from 40 eV to 10 eV in steps of 10 eV and its results are summarized in Fig.\ref{figure2}, where the different curves in each frame correspond to different temperatures (blue, green, red, black in order of decreasing temperature).

From the upper left frame it is clear that lower temperatures are associated with flatter and broader density profiles, so that the particles are transported deeper in the SOL, towards the walls of the machine. In particular, if we use the minimum of $\lambda_n$ close to the separatrix as a reference for layer thickness, the SOL width goes from 2.5 cm at 40 eV to 13 cm at 10 eV, roughly following an inverse linear dependence for the higher temperatures. This effect does not have a counterpart in the turbulence statistics, which remain roughly unchanged for all the temperatures. Although the upper right frame of Fig.\ref{figure2} displays minor variations in the PDFs, the overall behaviour of the fluctuations is consistent with the universality of the turbulence observed in TCV experiments \cite{Graves2005}   

Also in this case, the effective diffusion and velocity, shown in the lower frames of Fig.\ref{figure2}, vary with the radial position. In addition, their level depends on $T_0$ as well as their shape, which does not appear to be self-similar. This reinforces the notion \cite{Naulin2007} that SOL turbulence modelling using Fick's law is problematic, and can be done only with the support of nonlinear codes such as ESEL and with a case by case approach. 

Finally, it is useful to remark that lower edge temperature can be easily obtained by fuelling the plasma, increasing the line averaged density or reducing the heating power. Using this interpretation, the results we find are consistent with the experimental observations, i.e. a higher line averaged density corresponds to broader SOL profiles (because the edge temperature decreases). 

\subsection{Plasma current}

\begin{figure}[!h]
\centerline{\scalebox{0.8}{\includegraphics[clip]{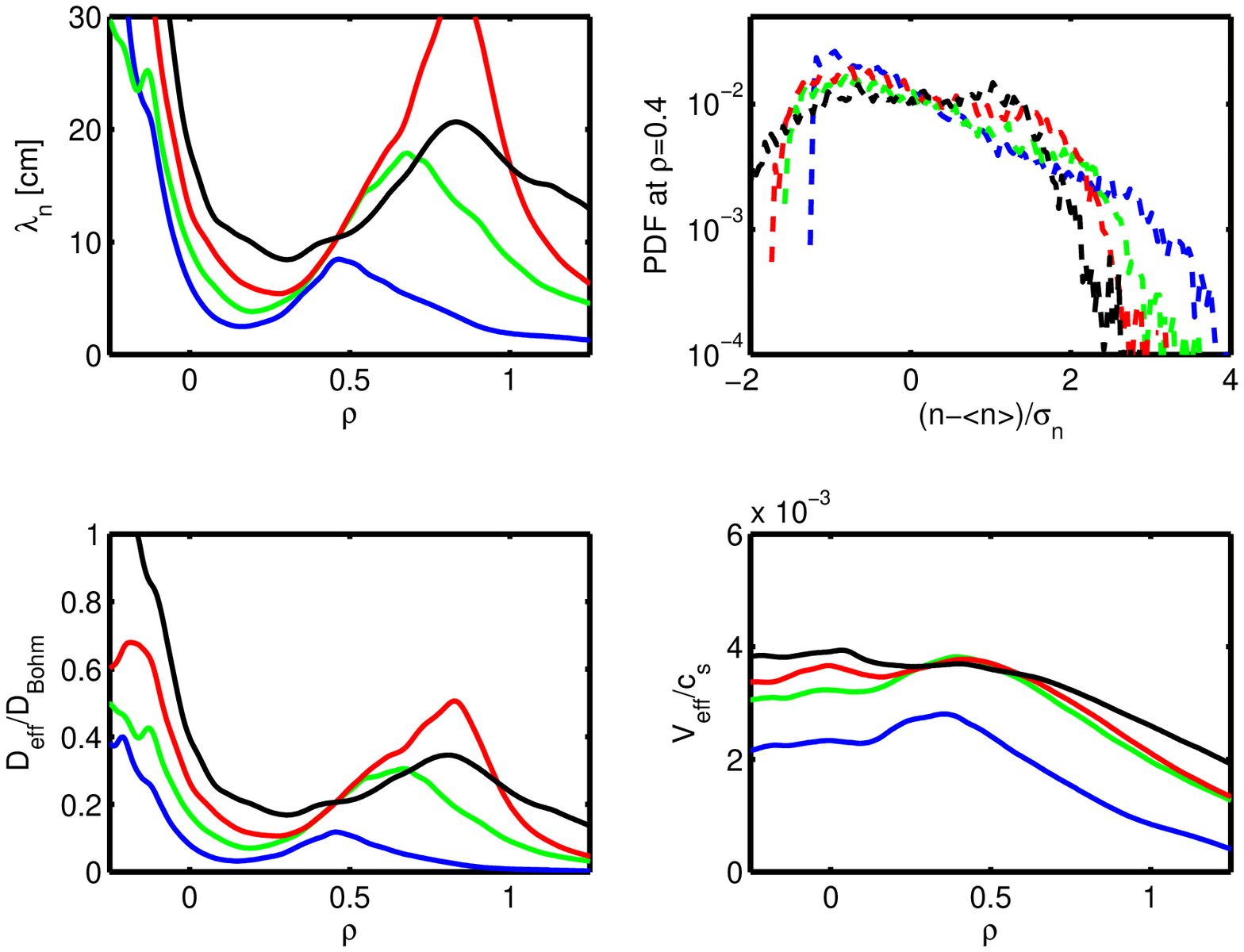}}}
\caption{}
\label{figure3}
\end{figure}

The parallel current, $I_p$ controls the magnetic configuration of the plasma. In particular, both the edge safety factor, $q\sim I_p^{-1}$ and the parallel connection length in the SOL, $L_\parallel\sim q$ are affected by its variation. This, in turn, modifies the neoclassical dissipation and the parallel loss terms and ultimately the properties of the SOL. In the model that we used, the plasma current does not directly appear. As a consequence, its effect was simulated by scanning $q$ and $L_\parallel$ at the same time, while keeping all the other parameters unchanged. In particular, we investigated the following cases: $q=7.11$ and $L_\parallel=10$ m (blue curves in the Fig.\ref{figure3}), $q=8.73$ and $L_\parallel=12.28$ m (green curves), $q=10$ and $L_\parallel=14.06$ m (red curves) and finally $q=11.5$ and $L_\parallel=16.17$ m (black curves). These high values of the safety factor are due to the particular magnetic geometry of MAST, and are related to its tight aspect ratio.  

As shown in Fig.\ref{figure3}, increasing the plasma current reduces the SOL density width. At the same time, the PDFs of the density fluctuations are similar to each other, although a trend toward less positive large events (blobs) can be identified when the current is decreased. It is interesting to note that the profile of the effective velocity seems to be relatively insensitive to variations at small currents, denoting a nonlinear dependence on $q$ and $L_\parallel$.

\subsection{Parallel connection length}

\begin{figure}[!h]
\centerline{\scalebox{0.8}{\includegraphics[clip]{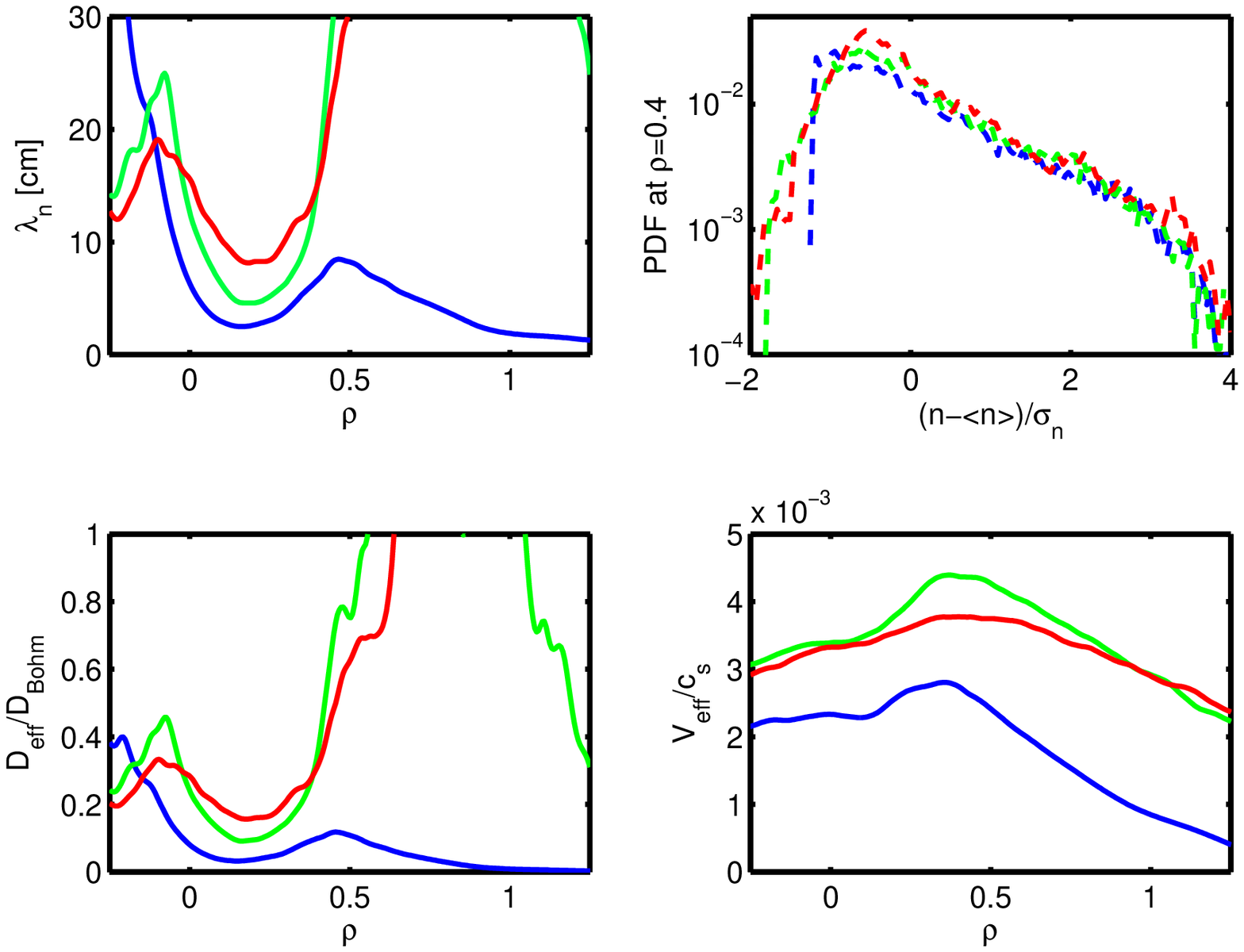}}}
\caption{}
\label{figure4}
\end{figure}

We complete our scans with the effect of the parallel connection length. This quantity is related to the typical length scale of the plasma filaments along the magnetic field and it determines the losses to the divertor plates. Investigating how this parameter affects the SOL is useful in the perspective of understanding the regime of operation of advanced exhaust solutions, such as the Super-X divertor \cite{Valanju2009} that will be installed on MAST-Upgrade. In particular, with this new design the outer midplane to divertor parallel distance will increase by a factor three so that the current 10 m will become 30 m. This is the length range that we describe in Fig.\ref{figure4}, where the blue curve is the reference case with $L_\parallel=10$ m, the green curve has $L_\parallel=20$ m and the red curve $L_\parallel=30$ m. 

The upper left frame in Fig.\ref{figure4} shows that the density decay length increases as the parallel connection length is increased, particularly in the far SOL. This profile broadening is due to the fact that the parallel losses are reduced since the filament must travel a longer distance to reach the divertor plates. On the other hand, the distribution of the turbulent fluctuations is again unaffected and reaches degrees of similarity analogous to those found in the edge density scan. 

The profile broadening in the far SOL corresponds to an increase of $D_{eff}$, which becomes extremely large for $\rho>0.5$. Both the effective diffusion and velocity saturate to a similar profile for large connection lengths, indicating a complicated functional dependence of these quantities with $L_\parallel$.       

\section{Effect on the particle fluxes}

\begin{figure}[!h]
\centerline{\scalebox{0.8}{\includegraphics[clip]{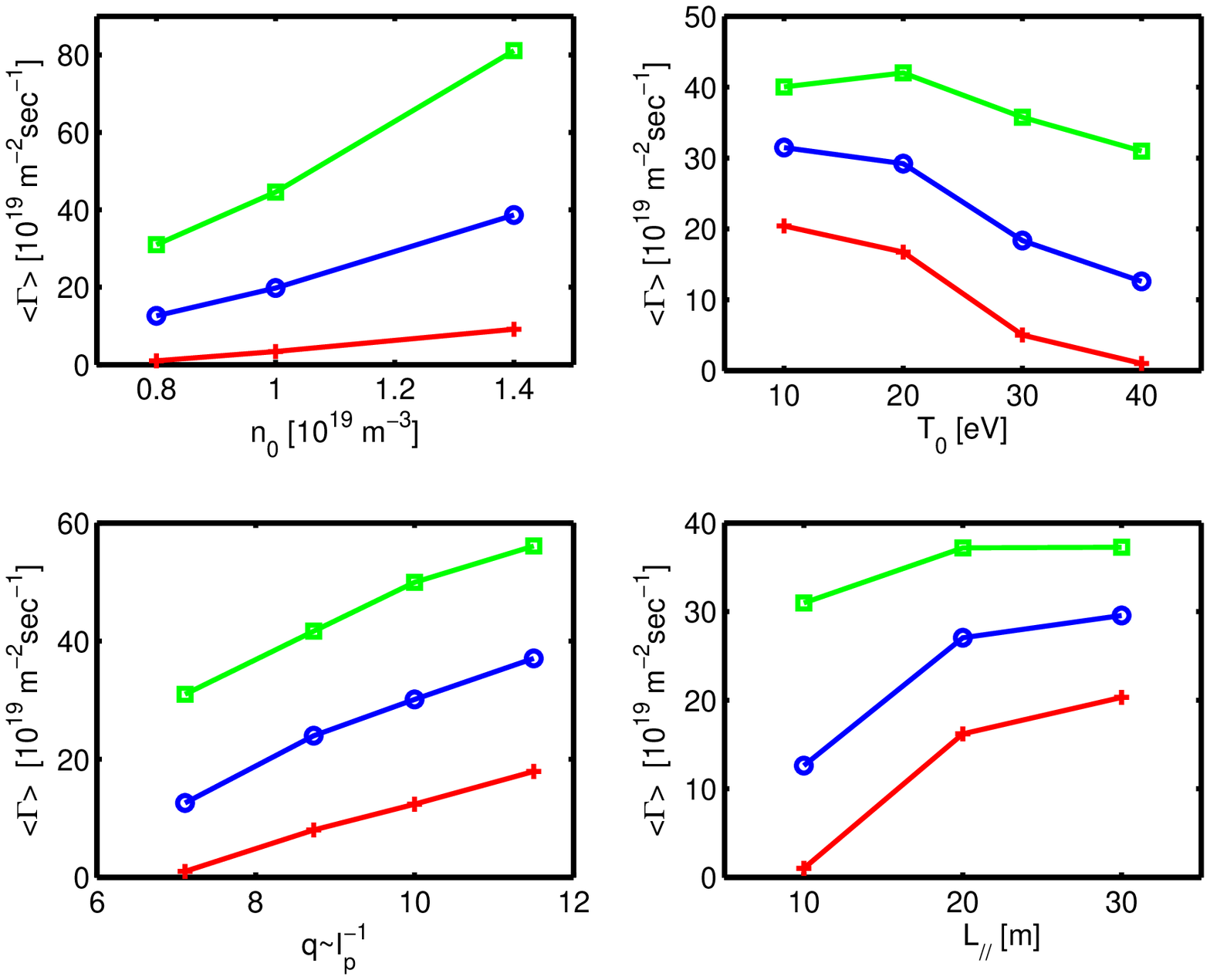}}}
\caption{}
\label{figure5}
\end{figure}

We now discuss how changes in edge quantities affect the cross field particle flux in the SOL, $<\Gamma>=<nV_r>$. Our results are summarized in Fig.\ref{figure5}, where the fluxes are evaluated at the last closed flux surface (green line) and at the radial positions $\rho=0.4$ (blue line) and $\rho=1$ (red line). We first notice that the edge density and plasma current scans show a roughly linear behaviour. This is more evident in the far SOL, where $<\Gamma>_{\rho=1}\approx 14 n_0$ and $<\Gamma>_{\rho=1}\approx 3.8 q$ (both these trends are consistent with experimental observations \cite{Garcia2007a,Garcia2007b}). The dependence on the edge temperature is more complicated and cannot be captured by simple scalings. 

Similarly, the scan in the connection length seems to suggest a saturation of the fluxes when the divertor leg becomes sufficiently long. This is due to the fact that, within our model, a longer connection length allows the blob to propagate radially almost unhindered, producing broad density and temperature profiles. At the same time, for higher $L_\parallel$ the filaments carry particles deeper in the plasma, so that at $L_\parallel=10$ m less than $10\%$ of the perpendicular flux at $\rho=0.4$ reaches $\rho=1$, while this ratio jumps to almost $70\%$ at $L_\parallel=30$ m. Conversely, for the $n_0$ scan $<\Gamma>_{\rho=1}/<\Gamma>_{\rho=0.4}$ remains between $0.1$ and $0.2$ for all the edge densities considered, as the loss terms do not depend on them.

\begin{figure}[!h]
\centerline{\scalebox{0.8}{\includegraphics[clip]{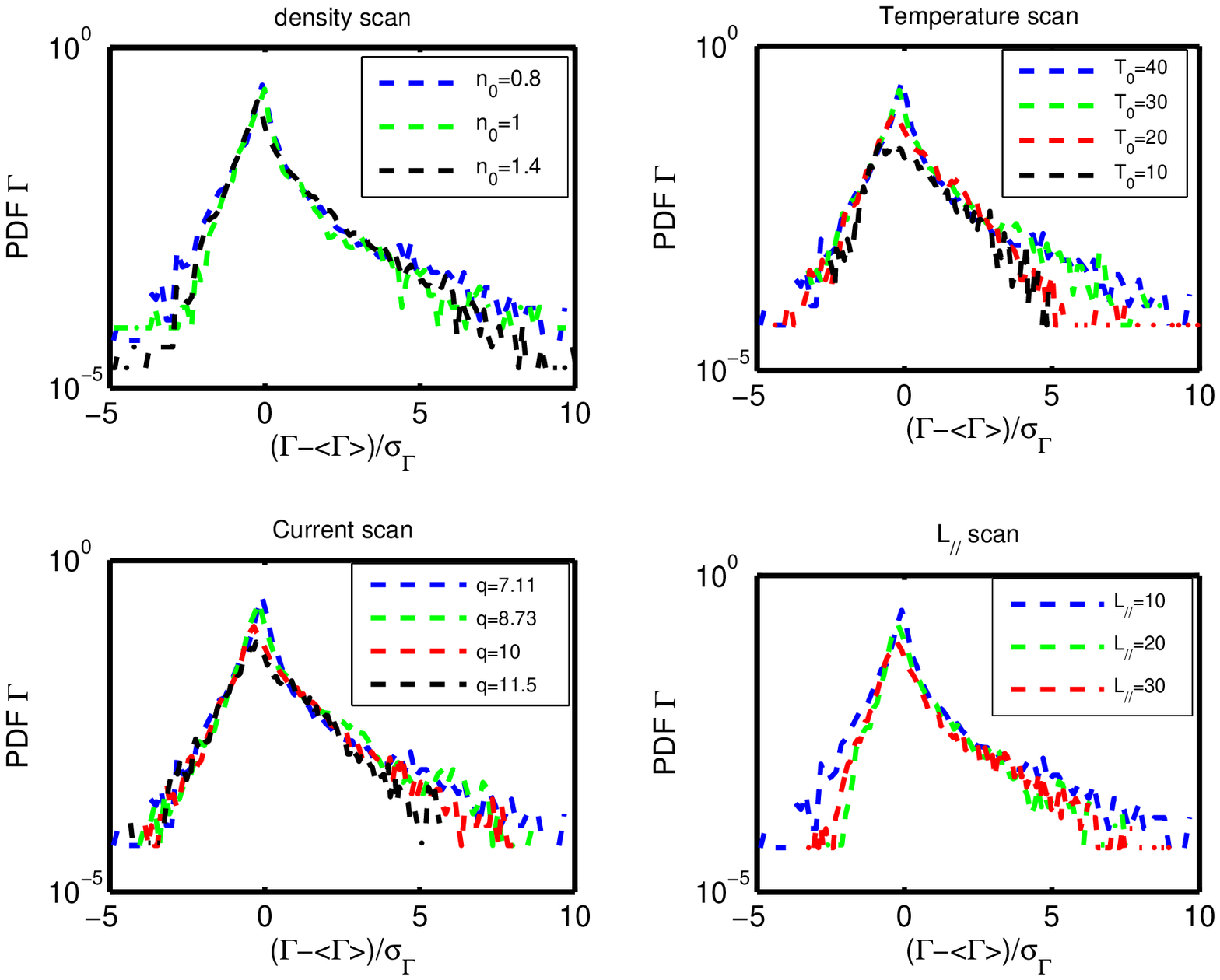}}}
\caption{}
\label{figure6}
\end{figure}

Finally, the statistics of the fluctuations of the particle flux are remarkably invariant. In previous studies \cite{Garcia2007a}, the experimental flux (local, not poloidally averaged) showed self-similarity for different line averaged density. We too find that this universality persists when the edge conditions change. In particular, Fig.\ref{figure6} shows that in the regime we investigated edge density, temperature, plasma current and connection length do not affect the PDF of the particle flux (which is evaluated at $\rho=0.7$). This suggests that the particle transport is dominated by a robust mechanism which is properly captured in our simulations.  

\section{Conclusions}

We carried out a numerical characterisation of the L-mode operating regime of the spherical tokamak MAST. Our approach allowed us to determine how specific edge quantities affect the particle transport in the SOL. We found that the statistics of the density fluctuations is not strongly affected by the plasma conditions inside the separatrix, while the density profile is. Specifically, the density profile in the SOL broadens when the current and the temperature decrease or when the connection length increases. The profiles seem to be unaffected by the edge density. This insight can help to better understand and design experimental scans, which typically affect more than one edge parameter at the same time. 

For each scan, we calculated the effective diffusivity and velocity, with the aim of identifying well defined trends. The strong radial dependence of these quantities, together with their sensitivity to the edge parameters does not allow a simple interpretation of the perpendicular particle transport in terms of Fick's law.

Finally, we related the perpendicular particle fluxes at three radial positions in the SOL to the plasma conditions inside the separatrix. We found that the fluxes in the far SOL depend linearly on both the edge density and current. The other edge quantities have more complicated effects on $<\Gamma>$, but they all have clear trends: the flux increases with higher $n_0$, lower $T_0$, lower $I_p$ and longer $L_{\parallel}$.         

The model used does not include distributed ionization sources or neutrals, and in general atomic physics. In addition, it is electrostatic and the parallel dynamics is described with \textit{ad hoc} terms, which might be modified by kinetic effects as well as by the magnetic shear. Drift wave physics, which is not captured in the equations, could also contribute to determine the filament penetration. Despite these limitations, our results are largely consistent with experimental observations. In particular, an approximate $1/I_p$ dependence for $\lambda_n$ is captured (details in \cite{Militello2012b,Militello2012c}) and, once the heat flux profiles are computed (not done in this paper, see \cite{Militello2012c}), trends compatible with recent experimental scalings \cite{Scarabosio2012} can be properly recovered (qualitatively and in some cases also quantitatively). 

\section{Acknowledgements}

This work was funded by the RCUK Energy Programme under grant EP/I501045 and the European Communities under the contract of Association between EURATOM and CCFE. The views and opinions expressed herein do not necessarily reflect those of the European Commission.

\pagebreak
\section*{Figure captions}
\noindent \textbf{Figure 1.} All the frames: scan in the density with $n_0=0.8\cdot 10^{19}$ $m^{-3}$ (blue curves), $n_0=1\cdot 10^{19}$ $m^{-3}$ (green curves) and $n_0=1.4\cdot 10^{19}$ $m^{-3}$ (black curves). Upper left frame: decay length of the time and poloidal averaged density profile as a function of the radial variable. Upper right frame: probability distribution function at $\rho=0.4$. Lower left frame: normalized effective diffusivity as a function of the radial variable. Lower right frame: normalized effective velocity as a function of the radial variable.\\

\section*{Figure captions}
\noindent \textbf{Figure 2.} All the frames: same as Fig.1 with scan in the temperature with $T_0=40$ eV (blue curves), $T_0=30$ eV (green curves), $T_0=20$ eV (red curves) and $T_0=10$ eV (black curves).\\

\section*{Figure captions}
\noindent \textbf{Figure 3.} All the frames: same as Fig.1 with scan in the current with $q=7.11$ and $L_\parallel=10$ m (blue curves), $q=8.73$ and $L_\parallel=12.28$ m (green curves), $q=10$ and $L_\parallel=14.06$ m (red curves) and $q=11.5$ and $L_\parallel=16.17$ m (black curves).\\

\section*{Figure captions}
\noindent \textbf{Figure 4.} All the frames: same as Fig.1 with scan in the connection length with $L_\parallel=10$ m (blue curves), $L_\parallel=20$ m (green curves) and $L_\parallel=30$ m (red curves).\\

\section*{Figure captions}
\noindent \textbf{Figure 5.} Effect of the scans of the edge parameters on the poloidally and time averaged radial particle flux at $\rho=0$ (green curves with square markers), $\rho=0.4$ (blue curves with circle markers) and $\rho=1$ (red lines with cross markers).\\

\section*{Figure captions}
\noindent \textbf{Figure 6.} Probability distribution functions of the perpendicular particle flux calculated at $\rho=0.7$ for all the edge parameter scans (note that the densities are expressed in units of $10^{19}$ $m^{-3}$, the temperatures in eV and $L_{\parallel}$ in m).\\

\pagebreak
\renewcommand{\baselinestretch}{1.0}

\pagebreak
\renewcommand{\baselinestretch}{1.0}

\pagebreak
\renewcommand{\baselinestretch}{1.0}

\pagebreak
\renewcommand{\baselinestretch}{1.0}

\pagebreak
\renewcommand{\baselinestretch}{1.0}

\end{document}